# Precision measurements with LPCTrap at GANIL

E. Liénard[1], G. Ban[1], C. Couratin[2], P. Delahaye[3], D. Durand[1], X. Fabian[1], B. Fabre[4], X. Fléchard[1], P. Finlay[2], F. Mauger[1], A. Méry[5], O. Naviliat-Cuncic[6], B. Pons[4], T. Porobic[2], G. Quéméner[1], N. Severijns[2], J.C. Thomas[3], Ph. Velten[2]

[1] *LPC CAEN, ENSICAEN, Université de Caen, CNRS/IN2P3, Caen, France*

[2] *Instituut voor Kern- en Stralingsfysica, KU Leuven, B-3001 Leuven, Belgium*

[3] *GANIL, CEA/DSM-CNRS/IN2P3, Caen, France*

[4] *CELIA, Université Bordeaux, CNRS, CEA, Talence, France*

[5] *CIMAP, CEA/CNRS/ENSICAEN, Université de Caen, Caen, France*

[6] *NSCL and Department of Physics and Astronomy, MSU, East-Lansing, MI, USA*

lienard@lpccaen.in2p3.fr

**Abstract**. The experimental achievements and the results obtained so far with the LPCTrap device installed at GANIL are presented. The apparatus is dedicated to the study of the weak interaction at low energy by means of precise measurements of the $\beta$–$\nu$ angular correlation parameter in nuclear $\beta$ decays. So far, the data collected with three isotopes have enabled to determine, for the first time, the charge state distributions of the recoiling ions, induced by shakeoff process. The analysis is presently refined to deduce the correlation parameters, with the potential of improving both the constraint deduced at low energy on exotic tensor currents ($^6\text{He}^{1+}$) and the precision on the $V_{ud}$ element of the quark-mixing matrix ($^{35}\text{Ar}^{1+}$ and $^{19}\text{Ne}^{1+}$) deduced from the mirror transitions dataset.

## 1 Introduction

Correlation measurements in nuclear β decays enable to probe the structure of the weak interaction, complementarily to high energy physics experiments [1,2]. In particular, the study of the angular correlation between the two leptons gives access to the parameter $a_{\beta\nu}$ sensitive to the existence of exotic currents, scalar or tensor, beyond the *V-A* structure of the Standard Model (SM). This parameter depends quadratically on the coupling constants associated to the different currents considered in the weak interaction. In the frame of the SM, for allowed nuclear decays it is given by:

$$a_{\beta\nu} = \frac{1-\rho^2/3}{1+\rho^2} \quad \text{where} \quad \rho = \frac{C_A M_{GT}}{C_V M_F}$$ is the mixing ratio of the transition, $C_A$

and $C_V$ are the coupling constants associated to the axial-vector and vector currents respectively, $M_F$ and $M_{GT}$ are the Fermi (*F*) and Gamow-Teller (*GT*) nuclear matrix elements. As a consequence, $a_{\beta\nu}$ equals 1 (-1/3) for a pure *F* (*GT*) transition. Any deviation from these values would imply either a departure from the allowed approximation or the presence of new physics beyond the SM. However, the distribution of events also depends on the Fierz term, *b*, which arises from the interference between exotic and standard currents and is therefore null in the SM. This particularity enables to consider that the effective parameter that is determined in a $\beta-\nu$ angular correlation experiment is actually [6]:

$$\tilde{a}_{\beta\nu} = a_{\beta\nu} / (1 + \langle b \frac{m}{E_e} \rangle)$$ where the brackets < > mean a weighted average over the

measured part of the $\beta$ spectrum, *m* and $E_e$ are respectively the mass and the total energy of the $\beta$ particle.

The first candidate studied with LPCTrap at GANIL was $^6$He [3], which is an ideal case to probe the tensor components. Until now only one experiment, performed in 1963, has reached a relative precision at the level of 1% (1σ) in this decay, yielding $a_{\beta\nu}$ = -0.3308(30) [4,5]. This result contributed significantly to fix the current limits on tensor contributions deduced from the correlation measurements performed at low energy [6].

For mirror transitions, the measurement of $a_{\beta\nu}$ also allows for a precise determination of the mixing ratio $\rho$. This parameter, combined with precise half-life, branching ratio and masses, can be used to compute the $V_{ud}$ element of the Cabibbo-Kobayashi-Maskawa quark-mixing matrix [7]:

$$V_{ud}^2 = \frac{4.794 \times 10^{-5}}{(Ft_{1/2}) G_F^2 |M_F|^2 (1 + \Delta_R)(1 + \frac{f_A}{f_V}\rho^2)} \quad (1)$$

where $Ft_{1/2}$ is the corrected *ft*-value of the transition, $G_F$ is the Fermi coupling constant, $\Delta_R$ is a transition-independent radiative correction and $f_A$ ($f_V$) is the statistical rate function computed for the *GT* (*F*) component. In the T=1/2 mirror transitions [8], the mixing ratio is always the least known parameter [9]. In the case of $^{35}$Ar and $^{19}$Ne, all parameters involved in the determination of $V_{ud}$, except $\rho$, are presently known with relative precisions below the $10^{-4}$ level. Accurate correlation measurements in their decays would then enable to significantly improve the current value of $V_{ud}$ deduced from the mirror transitions database [7]: $V_{ud} = 0.9719(17)$, which is a factor of 10 less precise than the result deduced from the pure *F* transitions [10]. $^{35}$Ar and $^{19}$Ne are two species also studied with LPCTrap [11].

In all experiments performed in the past, $a_{\beta\nu}$ was always deduced from the distribution of a kinematic parameter of the recoiling daughter ion (RI), since the $\nu$ detection is not efficient. Because of the very low kinetic energy of the RI, traps offer an ideal environment to confine the radioactive source and to ensure minimal disturbance for the RI motion [12-16]. The central element of LPCTrap is a transparent Paul trap [17], allowing the detection in coincidence of the $\beta$ particle and the RI. The setup is installed at the low energy beam line, LIRAT, of the GANIL/SPIRAL facility.

## 2. Performances of LPCTrap

At GANIL, the low energy radioactive beams delivered to LPCTrap are provided by the SPIRAL ECR source with a typical energy dispersion of 20 eV at 10 keV kinetic energy. A radio-frequency quadrupole cooler buncher (RFQCB) is then used to reduce the beam emittance and to produce ion bunches. The RFQCB is connected to the transparent Paul trap by a short line with dedicated beam optics and diagnostics. A telescope made of a double-sided silicon strip detector and a thick plastic scintillator is used to detect the $\beta$ particles while the RI's are detected in coincidence thanks to a micro-channel plates position sensitive system. The detectors are set in a back-to-back configuration, combining the highest statistics and the better sensitivity to a tensor component. The LPCTrap setup is described in detail in [11,14] and references therein. Some important features are summarized here:

- The trigger of an event is given by the detection of a particle in the $\beta$ telescope. Then many parameters are measured, such as the RI time-of-flight (ToF), the positions of the two particles, the trap RF and the timestamp of the decay in the measurement cycle. The set of distributions enables to reduce the background and to fix and control systematic parameters of the experiment during the off-line analysis.

- The current version of the setup contains a recoil ToF spectrometer which allows the separation of the different charge states of the RI, due to the shakeoff process. This spectrometer makes LPCTrap a unique setup to measure the charge state distribution of the RI after the $\beta$ decay of singly charged ions.

- Until now, a measurement cycle of 200 ms (100 ms in the first experiment [14]) was used: an ion bunch is extracted and sent to the Paul trap each 200 ms. Actually the ions are kept in the trap during 160 ms and then extracted to measure the background during the remaining 40 ms, which is thus controlled continuously during the experiment.

The performances of LPCTrap are summarized in Table 1, for the last experiments performed with $^6$He, $^{35}$Ar and $^{19}$Ne. The last column ($N_{coinc}$) gives the number of "good" coincidences accumulated in some days, which includes a complete detection of the two particles (energy and position) when the ion cloud in the trap is at equilibrium (a buffer gas is injected in the trap to cool the ions), and the subtraction of the remaining background. This number also depends on many parameters such as the geometrical detection efficiency ($4.5 \times 10^{-4}$), the emission anisotropy of the $\beta$s and the RI in the decay, the rate of recoil neutrals…The lower statistics for $^{19}$Ne is due to the long half-life and several technical problems on the beam production encountered during the experiment.

**Table 1** Performances of LPCTrap during the last experiments with $^6$He, $^{35}$Ar and $^{19}$Ne (see text for details). $I_{beam}$ is the beam intensity at the entrance of LPCTrap; the buffer gas is used in the RFQ and the measurement trap to cool the ions; $\varepsilon_{LPCTrap}$ is the transmission efficiency of LPCTrap.

| Beam (year of exp.) | $I_{beam}$ (pps) | Buffer gas | $\varepsilon_{LPCTrap}$ cycle=200ms | Trapped radio-active ions/cycle | $N_{coinc}$ |
|---|---|---|---|---|---|
| $^6$He$^{1+}$ (2010) | $1.5 \times 10^8$ | H$_2$ | $5 \times 10^{-4}$ | $1.5 \times 10^4$ | $1.2 \times 10^6$ |
| $^{35}$Ar$^{1+}$ (2012) | $3.5 \times 10^7$ | He | $4 \times 10^{-3}$ | $2.5 \times 10^4$ | $1.5 \times 10^6$ |
| $^{19}$Ne$^{1+}$ (2013) | $2.5 \times 10^8$ | He | $9 \times 10^{-4}$ | $4.5 \times 10^4$ | $1.3 \times 10^5$ |

## 3. Results

### 3.1 The charge state distributions

The RI spectrometer of LPCTrap has enabled to determine for the first time the experimental value of the shakeoff probability in the decay of $^{6}\text{He}^{1+}$ ions [18]. This system, with one single electron, is an ideal case to test the sudden approximation commonly used in the theoretical descriptions of shakeoff processes. Our measurements have validated this fundamental approximation, yielding an experimental shakeoff probability, $P_{SO} = 0.02339(36)$ given at $1\sigma$, in excellent agreement with its theoretical prediction, $P_{th} = 0.02322$.

In the case of $^{35}\text{Ar}^{1+}$, several electrons are involved, and such a system has revealed the importance of other processes such as the Auger emission [19]. The experimental charge state distribution of the resulting $^{35}\text{Cl}$ ions (Fig. 1) can indeed be reproduced by calculations only if single and multiple Auger decays, subsequent to inner-shell shakeoff, are explicitly taken into account. For $^{19}\text{Ne}^{1+}$, preliminary calculations predict a significantly lower effect due to Auger emission process, leading to lower yields for the highest charge states. Such a behavior is qualitatively observed in the ToF distribution of the $^{19}\text{F}$ ions (Fig. 2), which is more strongly dominated by the two first charge states than the $^{35}\text{Cl}$ ionic counterparts in the $^{35}\text{Ar}^{1+}$ decay. The detailed analysis of these data is currently in progress to determine accurately the experimental charge state distribution, including the neutrals, and to achieve a more constraining comparison with upgraded calculations including recoil effects and possible shakeup contributions.

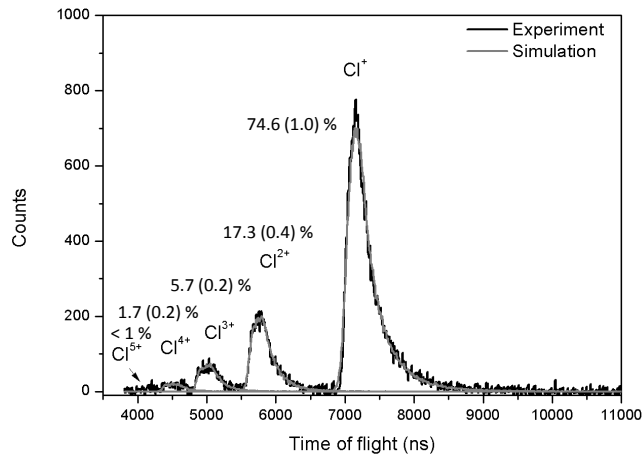

**Fig. 1**  ToF distribution of the $^{35}\text{Cl}$ RI resulting from $^{35}\text{Ar}^{1+}$ decay.

## 3.2 The $\beta$–$\nu$ angular correlation parameters

The extraction of $a_{\beta\nu}$ from the data requires realistic simulations of the experiments, containing a statistics significantly larger than the number of coincidences collected. The parameter is deduced from a fit of the RI ToF distribution using a linear combination of two distributions simulated with different values of $a_{\beta\nu}$ [14]. At least three parameters are left free in the fit: the value of $a_{\beta\nu}$, the total number of events and the distance, $d$, between the RI detector and the center of the Paul trap. Indeed, the last parameter is by far determined more precisely using the data themselves than a specific geometrical measurement. The two parameters, $a_{\beta\nu}$ and $d$, are correlated such that the minimization is completely unambiguous (see Fig. 13 in [14]).

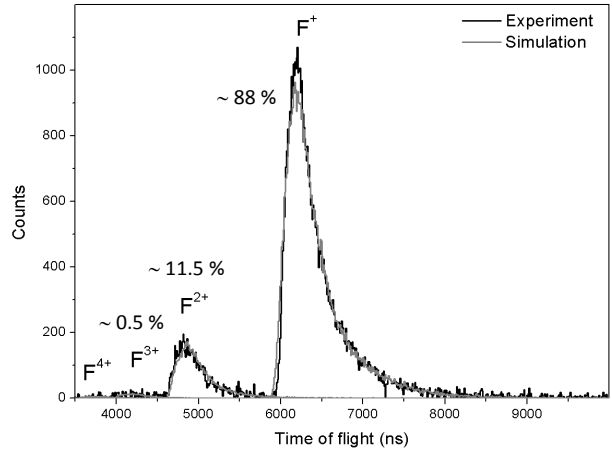

**Fig. 2**   ToF distribution of the $^{19}$F RI resulting from $^{19}$Ne$^{1+}$ decay.

For $^6$He, a first value at a relative precision of 3% (1$\sigma$) has been deduced from the data collected during a first experiment with limited statistics [14]:

$$a_{\beta\nu} = -0.3335\ (73)_{stat}\ (75)_{syst} \qquad (2)$$

The systematic error is largely dominated by the spatial distribution of the ion cloud inside the trap (91%), related to its temperature, and determined in an off-line experiment, using a $^6$Li source [20]. Simulations have shown that this parameter could also be determined with higher precision, if considered as a free parameter of the fit of the RI ToF distribution [21]. Indeed the shape of the leading edge of this distribution depends strongly on the cloud temperature and not on the value of $a_{\beta\nu}$. Finally all the dominant contributions to the final systematic error $(\Delta a)_{syst}$, except the $\beta$ scattering in the device, could be determined by the data themselves, with consequently a reduction of $(\Delta a)_{syst}$ with higher statistics. Concerning the $\beta$ scattering, the effect was estimated with GEANT4 simulations, leading to a relative contribution to $(\Delta a)_{syst}$ at the level of 0.6%, which remains still reasonable. To reduce it, new measurements should be considered in a large energy range, from 100 keV to some MeV, for which data on energy straggling and multiple scattering in thin materials are missing.

At higher statistics, the accuracy of the realistic simulations performed until now using different tools, mainly SIMION and GEANT4, is not yet sufficient to reproduce properly the whole set of distributions measured in the experiments. This analysis has revealed that the ion cloud simulation has to be improved. New simulation tools, using GPU technologies, are thus developed, aiming to describe the physical processes involved during the ion confinement in the trap (cooling, space charge effects, interaction in the RF field) in the most realistic manner [22]. The statistical precision expected in the three cases ($^6$He, $^{35}$Ar, $^{19}$Ne) is given in Table 2, according to $N_{coinc}$ (Table 1) and the precision obtained in the first experiment (eq. (2)). For $^{35}$Ar and $^{19}$Ne, the SM values for $a_{\beta\nu}$ given in the table were calculated following Ref. [8]. The mixing ratios were deduced from the ratios between the mirror $Ft$-values and the $Ft(0^+\to 0^+)$, including radiative corrections, but the extraction of the values of $a_{\beta\nu}$ does not include radiative corrections nor recoil effects. The current results are also given for comparison.

**Table 2** Statistical precision expected on $a_{\beta\nu}$ in the three nuclei studied at GANIL with LPCTrap (last column). SM and current values of $a_{\beta\nu}$ are given for comparison (see text for details).

| Isotope | SM value | Current value [Ref] | Projected 1σ precision |
|---------|----------|---------------------|------------------------|
| $^6$He | -0.3333 | -0.3308(30) [4,5] | 0.0015 |
| $^{35}$Ar | 0.9004[a](16) | 0.97(14) [23] | 0.0013 |
| $^{19}$Ne | 0.0438[b](8) | 0.00(8) [23] | 0.0046 |

[a]From Ref. [8]; [b]adapted from Ref. [8] with the new world average of $T_{1/2}$ [24]

## 4. Conclusion and outlook

In addition to the RI charge state distributions induced by the shakeoff process, LPCTrap has provided data with sufficient statistics to significantly improve the current values of $a_{\beta\nu}$ in the allowed $\beta$ decays of $^6$He, $^{35}$Ar and $^{19}$Ne. If the realistic simulations achieved with the advanced tools in development at LPC Caen become sufficiently accurate to reproduce the whole set of measured distributions, the systematic uncertainties should remain at a level of precision comparable to the statistical uncertainties.

For $^{35}$Ar, for example, the result would induce a significant gain (~ 1.7) on the $V_{ud}$ precision deduced from the study of mirror decays [7]. This perspective motivates future measurements at GANIL in mirror decays, using new beams which are presently under development [25], such as $^{33}$Cl and $^{37}$K. In these two cases, precisions similar to the $^{35}$Ar experiment are expected, considering an upgraded LPCTrap setup with increased detection efficiency, which is currently under investigation.

**Acknowledgements**   The authors thank the LPC staff for their strong involvement in the LPCTrap project and the GANIL staff for the preparation of the beams.

**Funding:** This work was supported in part by the Région Basse-Normandie, by the European networks EXOTRAP (contract ERBFMGECT980099), NIPNET (contract HPRI-CT-2001-50034) and EURONS/TRAPSPEC (contract R113-506065) and a PHC Tournesol (n° 31214UF).

**Conflict of Interest:** The authors declare that they have no conflict of interest.